put
 \documentstyle[12pt,a4wide]{article}
\catcode`@=11 
\@addtoreset{equation}{section} 
\def\@eqnnum{\hbox to .01pt{}\rlap{\rm \hskip -\displaywidth\theequation}}
\renewcommand{\theequation}{(\arabic{section}.\arabic{equation})}
\newenvironment{proof}{{\bf Proof:}}{}
\newtheorem{theorem}{Theorem}[section]
\newtheorem{lemma}[theorem]{Lemma}

\newtheorem{remark}{Remark}[section]

\newcommand{\qed}{$\bf \Box$}

\newcommand{\eps}{\varepsilon}
\newcommand{\be}{\begin{equation}}
\newcommand{\ee}{\end{equation}}
\newcommand{\p}{\partial}
\newcommand{\bea}{\begin{eqnarray}}
\newcommand{\eea}{\end{eqnarray}}
\newcommand{\nn}{\nonumber}

\begin{document}

\title{Fast and Slow solutions in General Relativity: The Initialization Procedure}

\author{Mirta S. Iriondo\thanks{Supported by STINT, The Swedish Foundation for International Cooperation in Research and Higher Education. }
 , Enzo O. Leguizam\'on
\thanks{Fellow of Se.CyT-UNC.}
\and and Oscar A. Reula
\thanks{Member of CONICET.}\\
{\small FaMAF, Medina Allende y Haya de la Torre,}\\
 {\small Ciudad Universitaria, 5000 C\'ordoba, Argentina}}

\maketitle
\begin{abstract}
We apply recent  results in the theory of PDE, specifically in problems
with two different time scales, on Einstein's equations near their
Newtonian limit.  The results imply a justification to Postnewtonian
approximations when  initialization procedures to different orders are
made on the initial data.  We determine up to what order initialization
is needed in order to detect the contribution to the quadrupole moment
due to the slow motion of a massive body as distinct from initial data
contributions to fast solutions and prove that such initialization is
compatible with the constraint equations. Using the results mentioned
the first Postnewtonian equations and their solutions in terms of Green
functions are presented in order to indicate how to proceed in
calculations with this approach.

\end{abstract}

\section{Introduction}
In recent papers there has been given a rigorous justification of the
Newtonian limit approximation in General Relativity. In one case,
\cite{reula-fri}, assuming symmetric hyperbolic equations for the
matter (including appropriate boundary conditions for it), and in the
other, \cite{rendall}, assuming Vlasov type matter, there has been
shown that given a Newtonian solution, there exists a nearby general
relativistic solution for a time intervall which is independent on the
limiting parameter.  The proofs of these results relay on ideas
pioneered by H-O Kreiss on dynamical systems with different time scales
which in turn relay on energy estimates for symmetric hyperbolic
systems.

The above mentioned results are based on a initialization procedure,
that is, the proximity of full relativistic solutions to the Newtonian
ones is obtained by choosing the initial data in a very special way,
basically ensuring that the time derivatives of the solutions at the
initial surface stay bounded on that limit.  This initialization
procedure is only needed to a finite order on the limiting parameter
(which is taken to be one over the speed of light) implying that fast
behaving parts of solutions stay under control order by order along
evolution.

In this paper we advance further into this problem by showing two
things: 

First that the General Relativity equations can be cast into a form on which
the standard theory of different time scales applies (c.f.
\cite{BK}, \cite{Maj}, and \cite{schochet}). This substantially improves the results in \cite{reula-fri}, for there there 
were singular terms outside the principal part which had to be dealt with 
in a very involved way.
This gives further information on
the behavior of the fast part of the solutions, that is the part that
comes from the failure to initialize data to all orders.  In particular, 
it gives the equations that the leading order fast behaving parts of
the solutions satisfies and so its bulk behavior. This result should be
important in studying the scattering of gravitational waves on slow
varying sources.

To be able to apply these standard results (estimates) we shall assume that
sources for Einstein's equations also satisfy symmetric hyperbolic
equations which are regular in the limit and, if boundary conditions
are needed to deal with them, that they are of such nature as to allow
for the estimates to hold. Admittedly there are some problems to grant
this to hold for some specific cases, but that is a problem on our
understanding of the description of normal matter, and are not very
much related to the dynamics of the gravitional degrees of freedom, so
we do not address this issue further.

Second that the initialization procedure can be done to higher
enough orders so that the quadrupole formula should follow. The
initialization procedure implies some relations between initial data
besides the one implied by the constraint equations, and so the above
mentioned results implies that showing that certain elliptic systems of
equations have solutions with the appropriate asymptotic behavior. The
result is only an argument, for the method used does not allow us to
have estimates, and so to control evolution, all the way up to some
portion of future null infinity, the region where the quadrupole
formula should hold. A rigorous result for that issue should follow by
studying evolution either along null cones, as considered by Winicour,
\cite{Winicour}, or along asymptotically null surfaces as studied by
Friedrich, \cite{Fried}.

The plan of the paper is as follows:

In the second Section we introduce the general theory of different
time scales systems and quote
the relevant Theorem.
We then follow \cite{reula-fri} and cast Einstein's
equations as a symmetric-hyperbolic-elliptic system. By assigning units
to the fields appearing in those equations one can determine how the
parameter $\eps=\frac1{c}$, where $c$ is the velocity of light, appears
in the system and in this way the solutions become a one parameter
family of solutions to Einstein's equations.  A further rescaling of
the fields is needed so that the structure of the equations is in the
form assumed in the hypotesis of the mentioned Theorem. 

In Section 3 we calculate the flux of energy coming from the slow part
of the solution (i.e the contribution to the flux of energy due to the
slow motion of a massive body) and the flux of energy coming from the
initial data without sources, that is the contribution from fast
solutions.  We conclude that one should initialize up to order three in
$\eps$ in order to isolate the contribution to the energy flux coming
from the slow part of the solution.  We further prove a Lemma stating
that such initialization is consistent with the constraint equations
and so that solutions with these characteristics exists.
 
In an Appendix we look at the solution to the initialized data 
to order $\eps$ and obtain an
explicit solution in terms of the shift vector $N^a$ and the source
$S^{ab}$. This is done in order to illustrate how the general setting
of the theory of different time scales produces the correct
Postnewtonian equations to that order and how to proceed if one whishes
to compute higher order corrections.


\section{Preliminaries}

The study of systems with different time scales reduces to the study of
systems of partial differential equations which are singular in the
limit of one of this time scales going to zero. The study of such a
limit distinguishes between different classes of solutions according to
their limiting behavior. Solutions which behave smoothly on the limit
are called slow, they move according to the time scale which remains
finite, those which do not have a well defined limit are called fast,
they have dependence on the time scale we are setting to zero. As an
example we consider the following system:  \bea u_t &=&
\frac{1}{\eps}(u_x - v_x), \\ v_t &=& v_x.  \eea Since the system is
linear we can consider its Fourier modes, with the ansatz

$$
 \left (
	      \begin{array}{l}
		   u\\
		    v\\ \end{array}
      \right )(x,t)= \left (
	      \begin{array}{l}
		   u_0\\
		    v_0 \end{array} \right )\mbox{e}^{i(kx-\omega t)},
$$ we obtain $$
 \left (
	      \begin{array}{l}
		   u\\
		    v\\ \end{array}
      \right )(x,t) = u_0\left (
	      \begin{array}{l}
		   1\\
		    0 \end{array} \right
      )\mbox{e}^{ik(x+\frac{t}{\eps})}+ v_0\left (
	      \begin{array}{l}
		   \frac{\eps}{1-\eps}\\
		    1 \end{array} \right )\mbox{e}^{ik(x+ t)}.  $$

It is clear that the first term in this expression is a fast solution, while the
second term is a slow one. Notice that for a solution to be slow it is
necessary that its time derivative stay bounded, in particular at
$t=0$. One of the main results of the theory of different time scales
is that, under some circunstances, this condition is also sufficient.
In particular this allow us to pick slow solutions from initial data
requiring that a number of consecutive time derivatives at the initial
time are bounded, this procedure is called initialization. Another
interesting output of the theory is that it gives precise information
on the evolution of the fast and slow parts.

We turn now to the general setting and quote the main results of the
theory in the following Theorem.  

\begin{theorem}[\cite{schochet1}, see also \cite{km2}]
\label{fast} Let \be \label{eq:s-h} A^0{}_{ij}(\eps u^k) \frac{\p}{\p
t} u^j =(\frac{1}{\eps}K^a{}_{ij} + A^a_1{}_{ij}(u^k,\eps) )\partial_a
u^j + B_{i}(u^k,\eps), \ee be a symmetric-hyperbolic system containing
the small parameter $\eps$; i.e., let the matrices $K^a$, $A^a_1{}$
and  $A^0{}$ be symmetric  and  $A^0{}$ positive definite. Assume that the
matrices $A^a_1{}$ and the vector $B$ are continuous in $\eps$
uniformly in $u$ for bounded $u$, and are $C^s$ in $u$ uniformly in
$\eps$ with $s\geq s_0\equiv [n/2]+2$ and that the initial data $u(0,x,\epsilon)=(q_{ab}(x,0),p_{ab}(x,0),r^{a}{}_{bc}(x,0))$ lies in $H^s({\bf
R}^n)$. Then the solution $u(t,x,\epsilon)$ exists for a
time $T$ independent of $\eps$. Moreover if the initial data for \ref{eq:s-h} is
$$
 u(0,x,\epsilon)=u_0(x)+\eps\;u_1(x)+{\cal O}(\eps^2)
$$
where
$$ K^a\partial_a u_0=0, $$  
then

$$ u(t,x,\epsilon)=u^0(t,x)+\epsilon
[v^0(t,x)+\tilde u (t,x,\epsilon)]+{\cal O}(\epsilon^2), $$ with $u^0$
and $\tilde u$ satisfying

\bea \label{u0} A^0(0){}u^0_{t} + A_{1}^a(u^0,0){}\partial_a u^0 + K^a
\partial_a v^0 & = & B_0 \nonumber \\
 K^a\partial_a u^0 & = & 0 \\
		 u^0(0,x) & = & u_{0}(x) \nonumber
\eea 
and 
\bea \label{u1} [A^0(0) + \eps u^0{} A^0_u(0)]\tilde u_{t} +
A_{1}^a(u^0,0) \partial_a \tilde u + \frac{1}{\eps} K^a \partial_a
\tilde u & = &  B_1  \nonumber \\
		\tilde u(0,x) & = & u_{1}(x) - v^0(0,x)
\eea
respectively

\bea
 B_0 & = & B(t,x,u^0,0) \nonumber \\ B_1 & = &
B_\eps{}(t,x,u^0,0)+(v^0+ \tilde u) B_u(t,x,u^0) \nonumber \\ & - &
(v^0 + \tilde u){}A^a_u(u^0,0) \partial_a u^0- A^0(0){}v^0_t -
A^a(u^0,0) \partial_a v^0, \nonumber \eea 
 and $v^0$   chosen in such a way that $K^a\partial_a
u^0=0 $ .  \end{theorem}

 The  solutions
of \ref{u1} are fast solutions depending on the choice of the
initial data. There exists further local (in ${\bf T}^n$) refinements of these results, that give us information about the behaviour of these fast solutions, see for example \cite{rauch1}, \cite{rauch2} and \cite{schochet}. These results should be an useful tool for the
understanding of problems like the scattering of gravitational waves by
slow moving bodies.

If  we would now want that  the fast solution appears in the second
order of $\eps$, $u^0$ and $u^1$  would have to satisfy equations
similar to \ref{u0} and  $\tilde{u}$ one similar to \ref{u1} (with
different sources), so that $u_t$ and $u_{tt}$ remain  bounded when
$\eps \to 0$ at $t=0$. Summarizing we can suppress the fast part up to
{\cal O}($\eps ^n$), demanding that $n$-time derivatives  stay bounded
when $\eps \to 0$.This procedure is called {\it initialization}, for
more details see \cite{schochet1} and \cite{BK}. Note that in the case
of General Relativity the choice of initial data for the initialization
procedure is not trivial for it must be consistent with the constraint
equations.

In order to give a formulation of the Postnewtonian limit in General
Relativity as an initialization procedure,   we use the variables given
in \cite{reula-fri}, such that the Einstein equations can be put as a
symmetric hyperbolic-elliptic  system.  Defining the lapse function as
\be
      N \equiv \frac{1}{1-\eps^2U}, \ee with $U$ being the Newtonian
potential and \be
     r^{ab}{}_c \equiv
		       \frac{1}{2\eps^3}
			(
			 \p_cq^{ab}-\frac12 q^{ab}q_{ed}\p_cq^{ed}
								  )
		\equiv
		       \frac{1}{2\eps^3 \sqrt q}\p_c
						    ( \sqrt q q^{ab}
								   )
\ee

\be
     p^{ab} \equiv
		   \frac{1}{\eps^2}\bar \pi^{ab},
\ee where $\p_c$ is the derivative operator associated to a flat
three-metric $e^{ab}$ and $q_{ab}$  is the three metric induced on the
hypersurfaces $ t=const.$.  The evolution equations become \bea
\label{eq:q.evol}
		 \dot q^{ab} & = & - 8\eps^2q^{ab}N \dot U
				   +2\eps^2 (1-4\eps^2 U)N^3
				   q^{\frac12}
				       (q^{ab} p - p^{ab}
								)\nonumber
								\\
				   &   &     - 2\eps^2 q^{\frac12}N
				   \bar D^{(a}N^{b)} +2\eps^2 q^{ab}
				\bar D_c N^c,
											   \eea

\bea \label{eq:p.evol}
		 \dot p^{ab} & = & - q^{\frac34} N^{\frac12}
		 \frac1{\eps}
				      (
				       q^{cd}\p_dr^{ab}{}_c -
				       2q^{c(a}\p_cr^{b)d}{}_d
								)
					- 2q^{\frac12}\frac1{\eps^2}
					q^{ab} (
				       \Delta U - \rho
						      )
								   \nonumber
								   \\
			     &   & + 2 S^{ab}
				   + \eps^2  N^c\p_cp^{ab} + \eps
				   F^{ab}
						(
						 \eps, r^{de}{}_c,
						 p^{de}, \p_c U , \p_c
						 N^d
								  )
\eea

\bea \label{eq:r.evol}
	    \dot r^{ab}{}_c & = &
	    -\frac{N^{\frac12}}{q^{\frac14}}\frac1{\eps}
				    (
				     \p_cp^{ab}
				     -2\delta^{(a}_c\p_dp^{b)d}
									  )
								\nonumber
								\\
			    &   & + \frac1{\eps}
				     \{
				       q^{ab} \p_c \p_d N^d - q^{d(b}
				       \p_c \p_d N^{a)}
								 \}
								 \nonumber
								 \\
			    &   & + \frac{4 N^{\frac12}}{q^{\frac14}}
				     \frac1{\eps} \delta^{(a}_cJ^{b)} +
				  \frac2{\eps} q^{ab} \p_c (N\dot U) +
				  \eps^2 N^d \p_d r^{ab}{}_c
								 \nonumber
								 \\
			    &   & - 2 \eps N q^{ab} N^d \p_c \p_d U
								 \nonumber
								 \\
			     &   & + \eps F^{ab}{}_c
						   (
						    \eps, r^{ab}{}_c,
						    p^{ab}, \p_c U ,
						    \dot{U}, \p_cN^a
							   )      ,
\eea
 where $N^a$ is the shift, $\bar D_a$ is the covariant derivative
 associated with $\bar q_{ab}$, $\Delta \equiv \bar q^{ab} \p_a \p_b$,
 $F^{ab}= F^{ab}(\eps, r^{ab}{}_c, p^{ab}, \p_c U , \p_cN^a )$, \\
$F^{ab}{}_c=F^{ab}{}_c(\eps, r^{ab}{}_c, p^{ab}, \p_c U , \dot
U,\p_cN^a)$, and all $F$'s in the equations are smooth pointwise
functions of all their arguments and the quadratic terms in
$r^{ab}{}_c$ and $p^{ab}$ are the highest powers the factors will
appear in and are ${\cal O}(\eps^2)$.

In these variables, the constraint equations become

\be \label{eq:const1}
		      \Delta U
 -\frac12 \eps \p_c r^{cd}{}_d = \rho
				 + \eps^2 F(\eps, r^{ab}{}_c, p^{ab},
				 \p_c U)
\ee

\be \label{const2}
	      - 2 \p_c p^{ca} = 4 J^a
				+ \eps^2 F^a(\eps, r^{ab}{}_c, p^{ab},
				\p_c U)
\ee

\be \label{eq:const11}
		  r^{ab}{}_c =  \frac1{2\eps^3 \sqrt q} \p_c (\sqrt q
		  q^{ab}).
\ee

The system \ref{eq:p.evol}-\ref{eq:r.evol} is  symmetric hyperbolic,
and has an energy estimate finite for fixed $\eps$ . In order to have
an energy estimate bounded in the limit when $\eps \to 0$ the authors
in \cite{reula-fri}, choose a gauge (i.e. a selection of lapse and
shift) that makes $r^{ab}{}_b \equiv 0$ and  $\dot r^{ab}{}_b = 0$. As
a matter of fact one could relax this gauge choice to $r^{ab}{}_b
\equiv {\cal O}(\eps^\beta)$ and  $\dot r^{ab}{}_b = {\cal
O}(\eps^\beta)$ where $\beta$ is a positive and arbitrary number that
guarantees that the energy estimates stay bounded even when $\eps \to
0$. In this paper we choose $\beta =3$, the reason of this choice
depends on the initialization procedure and it shall become clear in
the next Section.

 This choice implies the following equations for $N^a$:

\begin{equation} \label{elln} \partial_c D^c N^b - \partial^b D_c N^c -
4(J^b + \partial^b N\dot{U}) +2\eps^2 q^{cb}N^d\partial_c\partial_d
U-\eps^2G^b=0 \end{equation}

\be \label{divn}
	D_d N^d = -2 N \dot U, \ee combining equations \ref{elln} and
\ref{divn} we get

\be \label{elln1} \partial_c D^c N^b + \partial^b D_c N^c - 4 J^b
+2\eps^2 q^{cb}N^d\partial_c\partial_d U-\eps^2G^b=0 \ee and the
constraint equation \ref{eq:const1} becomes

\be \label{ellu}
			\Delta U= \rho
				 + \eps^2 F(\eps, r^{ab}{}_c, p^{ab},
				 \p_c U).
\ee

Thus the system given by \ref{eq:p.evol}, \ref{eq:r.evol},
\ref{const2}, \ref{eq:const11}, \ref{elln1} and \ref{ellu} constitutes
a symmetric hyperbolic-elliptic system, the hyperbolic  and elliptic
part are given by the equations \ref{eq:p.evol}, \ref{eq:r.evol} and
\ref{elln1}, \ref{ellu} respectively.

By inspection one can see   that the source $ B_{i}(u^k,\eps)$ in
\ref{eq:p.evol} and \ref{eq:r.evol} takes the  form  $
B_{i}(u^k,\eps)={C_{ij}\over \eps} u^{j}+ F_{i}(u^k,\eps)$, where
$C_{ij}$ is a constant matrix.  Since we want  to   apply  Theorem
\ref{fast}  on this system, the source has to be
smooth in $\epsilon$, thus  we  rescale the variables\footnote{A similar
rescaling is used in \cite{rendall}.} in the following way.  \be
\label{def:h} \bar r^{ab}_c= \eps r^{ab}_c\qquad \bar p^{ab}=\eps
p^{ab}\;\mbox{and}\; q^{ab}=e^{ab}+\eps^2\hat h^{ab}.  \ee Since
$F^{ab}$ and $F^{ab}_c$ have quadratic terms in the variables
$r^{ab}_c$ and $p^{ab}$ , we obtain a system such as \ref{eq:s-h}  with
a source term depending smoothly on $\eps$, namely:  \bea
\label{eq:q1.evol}
		 \dot q^{ab} & = & - 8\eps^2q^{ab}N \dot U
				   +2\eps (1-4\eps^2 U)N^3 q^{\frac12}
				       (q^{ab} \bar p - \bar p^{ab}
								)\nonumber
								\\
				   &   &     - 2\eps^2 q^{\frac12}N
				   \bar D^{(a}N^{b)} +2\eps^2 q^{ab}
				\bar D_c N^c,
											   \eea

\bea \label{eq:p1.evol}
		 \dot {\bar p}^{ab} & = & - q^{\frac34} N^{\frac12}
		 \frac1{\eps}
				      (
				       q^{cd}\p_d\bar r^{ab}{}_c -
				       2q^{c(a}\p_c\bar r^{b)d}{}_d
								)
					- 2q^{\frac12}\frac1{\eps}
					q^{ab} (
				       \Delta U - \rho
						      )
								   \nonumber
								   \\
			     &   & + 2 \eps S^{ab}
				   + \eps^2  N^c\p_c\bar p^{ab} + \eps
				   F^{ab}
						(
						 \eps, \bar r^{de}{}_c,
						 \bar p^{de}, \p_c U ,
						 \p_c N^d
								  )
\eea

\bea \label{eq:r1.evol}
	    \dot {\bar r}^{ab}{}_c & = &
	    -\frac{N^{\frac12}}{q^{\frac14}}\frac1{\eps}
				    (
				     \p_c\bar p^{ab}
				     -2\delta^{(a}_c\p_d\bar p^{b)d}
									  )+
								\nonumber
								\\
			    &   & + q^{ab} \p_c \p_d N^d
				       - q^{d(b} \p_c \p_d N^{a)}
								 \nonumber
								 \\
			    &   & + \frac{4 N^{\frac12}}{q^{\frac14}}
				     \delta^{(a}_cJ^{b)} + 2 q^{ab}
				  \p_c (N\dot U) + \eps^2 N^d \p_d \bar
				  r^{ab}{}_c
								 \nonumber
								 \\
			    &   & - 2 \eps^2 N q^{ab} N^d \p_c \p_d U
								 \nonumber
								 \\
			     &   & + \eps^2 F^{ab}{}_c
						   (
						    \eps, \bar
						    r^{ab}{}_c, \bar
						    p^{ab}, \p_c U ,
						    \dot{U}, \p_cN^a
							   )      .
\eea

In the same gauge as before, the constraint equations and the equation
for the shift vector become

\be
			 \Delta U= \rho
				 + \eps^2 F(\eps, \bar{r}^{ab}{}_c,
				 \bar{p}^{ab}, \p_c U)
\ee

\be
	      - 2 \p_c \bar{p}^{ca} = 4 \epsilon J^a
				+ \eps^2 F^a(\eps, \bar{r}^{ab}{}_c,
				\bar{p}^{ab}, \p_c U)
\ee

\be
		  \bar{r}^{ab}{}_c =  \frac1{2\eps^2 \sqrt q} \p_c
		  (\sqrt q q^{ab}),
\label{barr} \ee and \begin{equation} \partial_c D^c N^b + \partial^b
D_c N^c - 4J^b   + 3\eps F^b +\eps^2(4 Nq^{cb}N^d\partial_c\partial_dU-
F^{ab}{}_b)=0.  \label{eq:N} \end{equation}

From now on, to simplify the notation, we will work without bars on the
variables $p^{ab}$ and $r^{ab}{}_c$.
As stated in the Introduction we assume that the above system is part of a
bigger symmetric hyperbolic system which includes the equations for the
sources, they are assumed to be regular in $\eps$ and of such a type that
global estimates can be obtained.
 
The existence
of smooth one parameter family of solutions is guaranteed because they
prove that energy estimates corresponding to the hyperbolic part of the
system stay bounded when $\eps \to 0$ via a particular gauge choice,
the elliptic variables satisfy a G\aa{}rding estimate in terms of the
hyperbolic ones and the boundedness in time of the solutions are
guaranteed via the initialization procedure.

Another interesting  problem  which can be tackled with these tech\-niques
is whe\-ther for each given fast Postnewtonian
solution (i.e. a fixed slow background solution and a highly
oscillating part), there exists a solution to  the full Einstein
equations in the {\it same intervall of existence} and that it remains close
for that whole intervall to the first one in some norm. 
In other words, whether the solution to
\ref{eq:s-h} exists and converges to the solution of the limit equation
for as long as the General relativistic  solution exists.


\section{The Postnewtonian limit as an initialization procedure.}

The Theorem quoted above imply that solutions to Einstein's equations,
(the system given by \ref{eq:q1.evol} to  \ref{eq:r1.evol}),  can be
splitted into slow and fast ones, and this is ruled by the choice
of initial data.

In the context of General Relativity the slow part of the solution
contains information about the matter fields that are moving with
velocity much lower than $c$, that is with its own time scale, while
the fast part contains information essentially due to the arbitrariness
of initial data and so is not related directly to the sources.

In the context of symmetric-hyperbolic systems this means that on the
Postnewtonian approximations the highly oscillatory part (fast part) of
the solution should be suppressed up to higher order of $\eps$. For
example, if we initialize (choose initial data such that the time
derivative of the solution is bounded independently of $\eps$ at $t=0$)
up to order $\eps$ the slow part appears at zeroth order and the fast
one at first order.  The degree of initialization needed depends on the
problem at hand. In this Section we show that to get the quadrupole
formula one necessary to initialize up to order $\eps^3$.  This
guarantees that no contribution from fast solutions would be present in
the gravitational radiation flux to that order. We then show that such
initialization is possible.

To know at which order of $\eps$  we have to initialize, we estimate
the contribution to the  energy flux at ${\cal I}^+$ from both, sources
and vacuum (pure initial data) solutions. Since we are only interested
in the order of $\eps$ at which these contributions appear, it is
enough to estimate them in linearized gravity, for example see
\cite{Land-Lif}.

 To this end we keep the physical units in the Einstein equations (i.e
 $k=G\neq 1$), and we check the power at which  $\eps$ appears in the
 radiation due to the matter (quadrupole contribution, with zero
 initial data) and in the radiation due to the initial data (without
 sources).

In this approximation

$$ g_{\mu\nu}=\eta_{\mu\nu}+\delta g_{\mu\nu}, $$ and in the absence of
matter the  energy flux in the $x^1$ direction  due to the initial
data  can be calculated as

$$ ct^{01}=\frac1{32 \pi k \eps^3}\big [\delta \dot
g_{23}^2+\frac1{4}(\delta\dot g_{22}-\delta\dot g_{33})^2\big ], $$
where the dot means the time derivative. In the presence of matter,
with initial data $\delta g_{\mu\nu}=\delta\dot g_{\mu\nu}=0$, we can
use the same formula as above because at large distance from the bodies
the matter waves can be  considered as  plane waves.  By \ref{def:h}
$\delta  g_{ij}=\eps^2 \hat h_{ij}$, the flux of energy becomes $$
ct^{01}=\frac{\eps}{32 \pi k }\big [\dot {\hat
h}_{23}{}^2+\frac1{4}(\dot {\hat h}_{22}-\dot{\hat h}_{33})^2\big ], $$

Thus choosing the initial data $\hat h_{ij}= \eps^\alpha f_{ij}$, the
flux becomes $ct^{01}={\cal O}(\eps^{2\alpha+1})$, on the other side
the flux due to the matter is calculated as  $ ct^{01}={\cal
O}(\eps^5)$ (see the Appendix).  We conclude that initializing up to
order $\eps^3$  the fast part of the solution will contribute at a
higher order of $\eps$ to the flux. It means that the initial data will
be choosen  as $$ u|_{t=0}=u_0+\eps u_1+\eps^2 u_2+\eps^3 u_3+{\cal
O}(\eps^4).  $$

This is equivalent to demand that the first four time derivatives of
the solution stay bounded in $t=0$ when $\eps \to 0$ (c.f. \cite{BK}).

Besides an hyperbolic system Eintein's equations also include a
constraint system, that is, equations which relate the otherwise
arbitrary initial data fields between each other.  Thus it must be
chequed that these constraints are consistent with the initialization
procedure. To do that we add to the equations arising from the
initialization the constraint equations and show that in the resulting
system, the hierarchy of elliptic systems admits solutions with the
correct asymptotic fall off.

Defining the pseudotensor $h^{ab}$ \be
\sqrt{q}q^{ab}=\sqrt{e}e^{ab}+\eps^2h^{ab} \label{eq:hab} \ee and from
the definition of  ${ r}^{ab}{}_c$ given by \ref{barr}, we obtain $$
 r^{ab}{}_c =\frac{1}{2\sqrt q}\p_c(h^{ab}).  $$

Choosing  the gauge condition $r^{ab}{}_b=0$ and $\dot r^{ab}{}_b=0$ up
to order $\eps^3$ and demanding the boundedness of the first four time
derivatives we get that
 $p_i{}^{ab}$ and $h_i{}^{ab}$, $i=0,1,2,3$ should satisfy at $t=0$,
 the following differential equations:

\begin{eqnarray} \Delta_0 h_0{}^{ab}&=&0,\nn \\
\partial_cp_0{}^{ab}&=&0,\nn\\ \Delta_0 h_1{}^{ab}&=&0,\nn\\
\Delta_0(p_1{}^{ab}+\partial^{(a}N_0{}^{b)})&=&0,\nn\\
\Delta_0(p_2{}^{ab}+\partial^{(a}N_1{}^{b)})&=&0,\label{eqsystem}\\
\Delta_0 h_2{}^{ab}&=&2S_0{}^{ab}+F_0{}^{ab}+\partial^{(a}\dot
N_0{}^{b)}-e^{ab} F_0,\nn\\
\Delta_0(p_3{}^{ab}+\partial^{(a}N_2{}^{b)})&=&-2\dot S_0{}^{ab}-\dot
F_0{}^{ab}+2 e^{ab}\dot F_0- \partial^{(a}\ddot N_0{}^{b)}+ \nn\\ &&+
\partial^dF_0{}^{ab}{}_d- \partial^{(a}F_1{}^{b)}-2
e^{ab}\partial^e(N_0{}^d\partial_e\partial_dU_0),\nn\\ \Delta_0
h_3{}^{ab}&=&2S_1{}^{ab}-F_1{}^{ab}+\partial^{(a}\dot
N_1{}^{b)}-2e^{ab} F_1,\nn \end{eqnarray} where
$\Delta_0=e^{ab}\partial_a\partial_b$.  These equations are consistent
with the constraint up to order $\eps^3$, for they have been included
in the set.

\begin{lemma} The equation system for the initial data given by
\ref{eqsystem} admits a solution.  \label{lem:inidata} \end{lemma}
\begin{proof}
From the first five equations, we can choose
\be \label{eq:inidata} p_0{}^{ab}=0,\; r_0{}^{ab}{}_c=0, \;
r_1{}^{ab}{}_c=0,\; p_1{}^{ab}=-\partial^{(a}N_0{}^{b)} \ee and \be
\label{eq:inip2} p_2{}^{ab}=-\partial^{(a}N_1{}^{b)}.  \ee Therefore,
in order to guarantee  the existence of solutions, it is sufficient to
prove that the sources in the three last equations in \ref{eqsystem}
have the decay ${\cal O}(\frac{1}{r^3})$ (c.f.\cite{mac}).

Due to the  choice made in \ref{eq:inidata},  the functions
$F_0{}^{ab}$, $F_0$, $F_0{}^{ab}{}_c$, $F_1$, $F_1{}^a$ and
$F_1{}^{ab}$ are all linear functions of $\partial_a(U_0N_0{}^b)$ .
Similarly   $\dot F_0$ and $\dot F_0{}^{ab}$ depends linearly on
$\partial_t\partial_a(U_0N_0{}^b)$. But the vector fields $N_0{}^a$ and
$N_1{}^a$, and the function $U_0$ satisfy the following differential
equations:  \begin{eqnarray*}
\Delta_0N_0{}^a+\partial^a\partial_cN_0{}^c&=&4J_0{}^a,\\
\Delta_0N_1{}^a+\partial^a\partial_cN_1{}^c&=&4J_1{}^a,\\
\Delta_0U_0&=&\rho_0.  \end{eqnarray*}

Thus, because of the compactness of the support of the sources,  the
standard  theory of elliptic systems in ${\bf R}^n$ (c.f. \cite{mac})
assures that $U_0$,   $N_0{}^a$ and $N_1{}^a$, and  their time and
spatial derivatives decay as ${\cal O}(\frac{1}{r})$ and ${\cal
O}(\frac{1}{r^2})$ respectively. Hence  we conclude that the sources
decay fast enough as to ensure existence of solutions for $h_2{}^{ab}$,
$h_3{}^{ab}$, and $p_3{}^{ab}+\partial^{(a}N_2{}^{b)}$.

In order to have a well defined $p_3{}^{ab}$, it remains to prove that
$N_2{} ^a$ exists. Note that the gauge condition implies that $N_2{}
^a$ must satisfy

$$
\Delta_0N_2{}^a+\partial^a\partial_cN_2{}^c=4J_2{}^a-3F_1{}^a-4N_0{}^d\partial_d\partial^aU_0+2F_0{}^{ab}{}_b,
$$ and that the source in this equation has a fast enough  decay.
Thereby the existence of $N_2{}^{a}$ and so of $p_3{}^{ab}$ is
assured.\qed \end{proof}

This Lemma proves that we can initialize at least up to order $\eps^3$,
hence by  Theorem \ref{fast} we can assure that the fast solution appears
in a higher order than $\eps^3$

\section*{Appendix: The evolution of the initialized data}

In order to exemplify the initialization procedure,  we initialize up
to order $\eps$ and solve explicitly the equation for $u^0$, $u^1$ and
$v^0$, choosing the initial data $u_0$ and  $u_1$ as in
\ref{eq:inidata}. In fact  this choice gives that $\tilde u$ becomes a
slow solution suppressing the fast one to a higher order of
$\epsilon$.

\vspace{.2in} \noindent{\bf Claim:} \label{radiation} {\it Given the
initial data

\begin{eqnarray*}
 u(0,x)&=&\eps\left (
	      \begin{array}{l}
		   {p}_1^{ab}\\
		    {r}_{1c}{}^{ab}{} \end{array}
      \right ) (x)+ {\cal O}(\eps^2)\\ &=&\eps\left (
	      \begin{array}{c}
		   -\partial^{(a}N^{b)}\\
		    0 \end{array}
      \right ) (x)+ {\cal O}(\eps^2) \end{eqnarray*} where $N^a(x) $ is
the solution of the elliptic equation \ref{N-0} for $t=0$,  the
solution of \ref{eq:p1.evol} and \ref{eq:r1.evol} can be written as

\begin{eqnarray*}
 u(t,x)&=&\eps\left (
	      \begin{array}{l}
		   {p}_1^{ab}\\
		    {r}_{1c}{}^{ab} \end{array}
      \right ) + {\cal O}(\eps^2)\\ &=&\eps\left (
	      \begin{array}{l}
		   \tilde{p}^{ab}+v^{ab}\\
		    \tilde{r}^{ab}{}_{c}+v^{ab}{}_{c} \end{array}
      \right )+ {\cal O}(\eps^2), \end{eqnarray*} with \[v=\left (
	      \begin{array}{c}
		  -\partial^{(a}N^{b)}\\
		    0 \end{array} \right )\] and \begin{eqnarray*}
\tilde{p}^{ab}&=&  \int_{|{\bf x}-{\bf
x}'|\leq\frac{t}{\eps}}\frac{\Gamma_1^{ab}(t-\eps|{\bf x}-{\bf
x}'|,{\bf x}')}{|{\bf x}-{\bf x}'|}{\rm
d}^3x'+\partial^{(a}N^{b)}\nonumber\\ \tilde {r}_{c}^{ab}{}&=&\eps
\partial_c\int_{|{\bf x}-{\bf
x}'|\leq\frac{t}{\eps}}\frac{\Gamma_2^{ab}(t-\eps|{\bf x}-{\bf
x}'|,{\bf x}')}{|{\bf x}-{\bf x}'|}{\rm d}^3x', \end{eqnarray*} where $
\Gamma_1^{ab}=   -(\Delta\partial^{(a} N^{b)}+ 2\eps^2\dot S^{ab})$ and
$ \Gamma_2^{ab}=\partial^{(a}\dot N^{b)}+2S^{ab}$,  $\rho$ and $ J^a$
have compact support in $R^3$.  Furthermore, the asymptotic behavior of
$\tilde{u}$  is ${\cal O}\big (\frac{1}{r}\big )$.  }

\begin{proof} Since  $u_0=0$ and thereby  $K^a\partial_a u_0=0$,
Theorem \ref{fast} assures that $$ u(t,x,\epsilon)=u^0(t,x)+\epsilon
[v(t,x)+\tilde u(t,x,\epsilon)]+{\cal O}(\epsilon^2).  $$

By applying \ref{u0} and \ref{u1} on the equations system given by
\ref{eq:p1.evol} and \ref{eq:r1.evol}, we obtain that $u^0,$ and $v$  will satisfy

\bea \label{relu0} \dot {p}_0^{ab}+\partial^c{v}^{ab}{}_c -
2\partial^{(a}v^{b)d}{}_d&=& 0\nonumber\\
\dot{r}_{0c}{}^{ab}+\partial_c{} v^{ab} -2\delta_c^{(a}\partial_d
v^{b)d}&=&   -\partial _c\partial^{(b}
N^{a)}+4\delta_c^{(a}J^{b)},\nonumber\\ K^a\partial_au^0&=&0.  \eea

At this order of $\eps$ the constraint equations and the gauge become

\bea
			 \Delta U&=& \rho\nonumber\\
	       - 2 \p_c {p}_0^{ca} &=& 0\nonumber\\
		  {r}_{0c}{}^{ab}&=&  \frac1{2} \p_c
		  h_0^{ab}\nonumber\\
\partial_c \partial^c N^b + \partial^b \partial_c N^c&=& 4J^b,
\label{N-0} \eea where ${h}^{ab}$ is a pseudotensor defined in
\ref{eq:hab}.  Choosing

\[ v=\left (
	      \begin{array}{c}
		   -\partial^{(a} N^{b)}\\
		    0 \end{array}
      \right ), \] the system \ref{relu0} becomes \bea \label{relu01}
u^0_{t}  & = & 0  \nonumber \\
 K^a\partial_a u^0 & = & 0 \\
		 u_0(x) & = & 0 ,\nonumber
\eea which is trivially satisfied by  $u^0(t,x) = 0$.

Considering now the equation for $\tilde u$, namely

\bea \label{relu1} \tilde u_{t} +\frac{1}{\eps} K^a \partial_a \tilde u
&=&  B_1 \\ \tilde{u}(0,x)=u_1(x)-v^0(0,x)=0\nonumber \eea with \[
B_1=\left (
	      \begin{array}{c}
		   2{}S^{ab}+\partial^{(a} \dot{N}^{b)}\\
		    0 \end{array}
      \right ), \] we obtain the following system \bea \label{sis} \dot
{\tilde{p}}^{ab}+\frac{1}{\eps} \partial^c\tilde{r}^{ab}{}_c & = &
2S^{ab}+\partial^{(a} \dot N^{b)}  \\
\dot{\tilde{r}}^{ab}{}_c+\frac{1}{\eps} \partial_c{}\tilde{p}^{ab} & =
0 \eea and  the constraint and the gauge become

\bea
			 \Delta U&=& \rho\nonumber\\
	       \p_c {p}_1^{ca} &=& -2 J^a\nonumber\\
		  {r}_{1c}{}^{ab}&=&  \frac1{2} \p_c
		  h_1^{ab}\nonumber\\
\partial_c \partial^c N^b + \partial^b \partial_c N^c&=& 4J^b.
\label{gauge1} \eea

Since $\tilde{p}^{ab}={p}_1^{ab}-v^{ab}$ and
$\tilde{r}_{c}{}^{ab}=r_{1c}{}^{ab}$, instead of \ref{sis} we  consider
the differential equations \bea \label{sis1} \dot
{{p}}_1^{ab}+\frac{1}{\eps} \partial^c{r}_{1c}{}^{ab}{}& = &
2S^{ab}\nonumber  \\ \dot{{r}}_{1c}{}^{ab}+\frac{1}{\eps}
\partial_c{}{p}_1^{ab} & =& -\frac{1}{\eps} \partial_c{}\partial^{(a}
\dot N^{b)}.  \eea

The equations system \ref{sis1} is clearly consistent with the gauge
\ref{gauge1}. Taking the time derivative to the system above, we obtain

\bea \label{eq:onda1} \Box {p}_1^{ab}&=&-(\Delta\partial^{(a} N^{b)}+2
\eps^2\dot S^{ab})\nonumber\\ \Box {h}_1^{ab}&=&2\eps
(\partial^{(a}\dot N^{b)}+2 S^{ab})\nonumber\\ &:=& 2\eps\Gamma_2^{ab}
\eea
and the solution of the second equation in \ref{eq:onda1} can be
written as $$ {h}_1^{ab}=2\eps\int_{|{\bf x}-{\bf
x}'|\leq\frac{t}{\eps}} \frac{\Gamma_2^{ab}(t-\eps|{\bf x}-{\bf
x}'|,{\bf x}')}{|{\bf x}-{\bf x}'|}{\rm d}^3x'.  $$

Because of the decay of the source, when $|{\bf x}|
>\hspace{-.05in}>|{\bf x}'|$, we  write $|{\bf x}-{\bf x}'|
\approx|{\bf x}|- \hat n\cdot{\bf x}'$, where $\hat n=\frac{ {\bf
x}}{|{\bf x}|}$. Hence in this first approximation we have $$
{h}_1^{ab}\approx \frac{2\eps}{r}\int_{|{\bf x}-{\bf x}'|\leq
\frac{t}{\eps}} \Gamma_2^{ab}(t-\eps r+\eps \hat n\cdot{\bf x}',{\bf
x}'){\rm d}^3x', $$ similarly we calculate $p_1^{ab}$.\qed \end{proof}

Following this procedure we can generate the solution that has as
initial data  \ref{eqsystem}  and  the functions $u^2(t,x)$ and
$u^3(t,x)$ obey similar equations than $u^0(t,x)$ and $u^1(t,x)$ do.

\begin{remark}
 In the calculus of the flux of energy due to the matter fields, we use
 the pseudo-tensor field calculated above instead of    the spatial
 components of  the perturbation of the metric. The difference between
 them can be calculated using equations \ref{def:h} and \ref{barr}, and
 because they have the same expansion in $\eps$  they can be considered
 as equal.  \end{remark}

 \end{document}